\documentstyle[aps,floats]{revtex}
\begin{document}
\twocolumn[
\title
{The Decuplet Revisited in $\chi$PT}
\author{M. K. Banerjee and J. Milana}
\address
{Department of Physics, University of Maryland\\ College Park,
Maryland 20742, USA}
\date{DOE/ER/40762--065, UMPP \#96--19,  August 21, 1995}
\maketitle

\begin{abstract}\widetext
The paper deals with two issues. First, we explore the quantitiative
importance of higher multiplets for properties of the $\Delta$
decuplet in chiral perturbation theory.   In particular,
it is found that the lowest order one--loop contributions from the
Roper octet to the decuplet masses and magnetic moments are substantial.
The relevance of these results to the chiral expansion in general is discussed.
The exact values of the magnetic moments depend upon delicate cancellations
involving ill--determined coupling constants.  Second, we present new relations
between the magnetic moments of the $\Delta$ decuplet that are independent
of all couplings.    They are exact at the order of the chiral
expansion used in this paper.
\end{abstract}
\vglue0.25in
]

\narrowtext
\section{Introduction}
The success of chiral perturbation theory ($\chi$PT)\cite{Wphilo}
for understanding properties of the pseudoscalar mesons is
now well established\cite{GassLeut}.  The approach is based on the
existence of a systematic expansion in derivatives of the
pion's field and the pion's mass,
whereby  $m_\pi$ divided by some large scale,
 generated by the theory itself and typically  $\sim 4 \pi f_\pi$, becomes
 the perturbative expansion parameter.    For the purely mesonic sector this
expansion is in fact quadratic in the pion's mass, so that even for the
$SU_f(3)$ generalization,  $(m_K/4 \pi f_\pi)^2$ is still a
reasonablely small parameter.

The application to the baryon sector has however from the outset been
confronted with a variety of difficulties~\cite{GassNuc}.
For example, how to handle the nucleon's mass was
a problem only relatively recently solved~\cite{Wcount,JenkMan1}.
One unavoidable complication when including baryons is that
the chiral expansion is itself more complicated~\cite{GassNuc,Wcount}
than in the purely mesonic case,
and the expansion parameter is now only linear in $m_\pi$ ($m_K$).
A second pertinent complication involves the issue of resonances.
\footnote{Because of the large mass difference between the
$\rho$ and the pion, as
well as the quadratic nature of the chiral expansion,
this issue does not arise when
considering the sector of purely pseudoscalar, goldstone bosons.
The effects of the $\rho$ and other
such states are incorporated as part of the needed
input coupling constants of the theory.}

Originally it was conjectured~\cite{Wcount} that all such resonances
(and most notably the $\Delta$) need not be included as an explicit
degree of freedom,
{\it i.e.} that they could be ``integrated out''.
While at least one  notably active group~\cite{BerKaisM}
have maintained this viewpoint,
\footnote{Arguing that for a very limited region near
threshold in the $SU_f(2)$ defined theory,
the $\Delta$-nucleon mass difference can be considered ``large''}
 most researchers have subsequently found it untenable~\cite{Kolck}.
The $\Delta$ degree of freedom was first introduced into $\chi$PT
by Jenkins and Manohar in Ref.~\cite{JenkMan2}.
Recently, the importance of the $\Lambda^*(1405)$ for
understanding threshold kaon--nucleon
scattering lengths has also been realized~\cite{MRho1,Savage,MRho2}.

In this paper we discuss the role of higher--multiplets
for the properties of the
$\Delta$ decuplet at the one--loop level in $\chi$PT.
We consider the ${\cal O}(p^3)$ correction to the decuplet masses
and the  ${\cal O}(p^2)$, one--loop correction to the magnetic moments of the
decuplet (the electromagnetic vertex has chiral power $-1$ excluding
whatever power maybe assigned to electric charge).
Our criterion for which multiplets to consider
is that the average mass--splitting, $\delta_h$, between the multiplet and the
$\Delta$ decuplet be less than the mass of the kaon,
\footnote{$m_\eta$ is obtained at this order in $\chi$PT using
GMOR~\cite{GMOR}.}
\begin{equation}
\delta_h = \mid M_H - M_{10} \mid < m_K.\label{criteria}
\end{equation}
This criterion is based on the fact that when it is met, an expansion
in $m_K/\delta_h$ is not justified so that loop effects involving these
higher--mulitplet members as intermediate  states {\it cannot} be absorbed into
higher order terms in the chiral lagrangian.
Such loop effects place a fundamental limitation on any formulation of
$\chi$PT that omits the higher--multiplet as an explicit degree of freedom,
{\it even if the chiral expansion was then executed to all orders.}

By the convention of Ref. \cite{ourselves},
note that the chiral power of all $\delta$ is $1$.

For the case of the nucleon octet Eq. (\ref{criteria}) is clearly met,
\begin{equation}
\delta_N = M_{10} - M_8 = 226{\rm MeV},\label{deltaN}
\end{equation}
($M_{10} = 1377$MeV and $M_8 = 1151$MeV)
and is indeed the driving phenomenological reason for expecting that
the $\Delta$ cannot
be ignored for descriptions of the nucleon.

We focus here on the properties of the
$\Delta$ decuplet as opposed to those of the nucleon octet because of the
simple
reason that the mass splitting $\delta_h$ satisfies the criterion specified by
Eq.~(\ref{criteria}) for at least two resonances.

We will show that the Roper has a nontrivial effect on both decuplet
mass splittings
and magnetic moments.   We should also note that these results lead to  a
more general statement. When adding a loop  one must also add resonances
whose masses are within a kaon mass of the resonances already included.

The rest of this paper is organized thusly.  In Section (II) we
enumerate the various multiplets considered and their interactions
(and experimentally obtained couplings) with the
$\Delta$ decuplet utilizing the heavy baryon formalism
of Jenkins and Manohar~\cite{JenkMan1}.
In Section (III) we present the one--loop, $O(p^3)$ contributions to
the decuplet masses focussing on the violation to the
Decuplet Equal Spacing rule~\cite{nobel},
the sole quantity for which $\chi$PT makes a prediction at this order
in the chiral expansion~\cite{Jenkins,ourselves}.
In Section(IV) we consider the
one--loop,  ${\cal O}(p^2)$ results for the magnetic moments of the decuplet,
a subject first discussed in  $\chi$PT in
Ref.~\cite{bss} where though adherence
to the order of the chiral expansion was not strictly maintained.
We   demonstrate later that  strict adherence
is crucial for renormalizability.   The focus in both
Sections (III) and (IV) is the quantitative importance of the
higher--multiplets.
In addition new relations for the magnetic moments at this order in
$\chi$PT are derived that are independent of the intermediate multiplets
considered.   Their violation would be a clear measure of the
importance of higher chiral power terms in the expansion.
In Section (V) we conclude with a
discussion of the consequence of these results to the
loop--expansion in general in $\chi$PT.

\section{Higher Multiplets: Definitions and Couplings}
A number of multiplets
\footnote{See {\it e.g.} Table 30.4 in the Particle Data Group~\cite{PDG}.
As we are at  present only concerned with the average coupling of these
multiplets with the $\Delta$ decuplet, we ignore potentially
interesting questions as to the exact
$SU_f(3)$ composition of any particular excited state.~\cite{yasuo}
We also omit from consideration possible exotics.}
 satisfy the criteria Eq.~(\ref{criteria}).  Fortunately most of
these can be eliminated due to symmetry constraints.  For example,
flavor singlets such as the
$\Lambda^*(1405)$ do not couple to a decuplet via an $SU_f(3)$ octet
(the goldstone bosons).
Only slightly less straightfoward, a $1/2^-$ octet ({\it e.g.} the
N(1535) multiplet)
couples only through the lower components of the baryon spinors,
\begin{equation}
{\cal L}^i \sim \overline{T}^\mu \gamma_5 A_\mu B^*,
\end{equation}
which  vanishes to lowest order in the
heavy baryon expansion~\cite{JenkMan1}.
(Such states would, in principle, need to be considered in
higher--order calculations.)
Coupling to the $5/2^-$ octet likewise vanishes at lowest order.
With these eliminations,
we obtain that only octets or decuplets of baryons with quantum numbers
$1/2^+, \, 3/2^+, \, 3/2^-$ and $5/2^+$ need be considered.

The most important $1/2^+$ multiplet (beyond of course the nucleon's) is the
octet containing the Roper, N(1440).   A slight difficulty
arises in determining
\begin{equation}
\delta_R = M_8(1440) - M_{10}
\end{equation}
because one member of the Roper multiplet, the excited Cascade,
has not yet been  identified.  To get a reasonable approximate
value for its mass, we  use the
corresponding GMO relation~\cite{GMO}
\begin{equation}
M_{\Xi^*}=\frac32\,M_{\Lambda^*}(1600) +
\frac12\,M_{\Sigma^*}(1660) - M_{N^*}(1440),
\end{equation}
by which one estimates that $M_{\Xi^*} = 1790$.
The average value of the Roper multiplet is  thereby
\begin{eqnarray}
M_{8^*} &=& \frac18 (2 M_{\Xi^*} +  3 M_{\Sigma^*} +
M_{\Lambda^*} +2 M_{N^*})\nonumber\\
 &=& 1630 \, {\rm Mev},
\end{eqnarray}
from which one gets that
\begin{equation}
\delta_R = 253{\rm MeV}.
\end{equation}
For the $\Delta N^* \pi$ interaction one has
\begin{equation}
{\cal L}^i = {\tilde C} (\overline{T}^\mu A_\mu B^* + {\rm h. c.})
\end{equation}
in complete analogy to the leading $\Delta N\pi$ interaction in the
heavy fermion limit.~\cite{JenkMan2}   The coupling ${\tilde C}^2$ can
be obtained from
the observed decay of the $N^*(1440) \rightarrow \Delta \pi$ which has
a partial width
$\Gamma_{N^* \rightarrow \Delta \pi} \approx 90$MeV.
Comparing this with the decay of the $\Delta$,
$\Gamma_{\Delta \rightarrow N \pi} \approx 120$MeV one obtains that
\begin{equation}
{\tilde C}^2 \approx \frac12 {\cal C}^2.\label{couplingt}
\end{equation}
In principle, one other $1/2^+$ multiplet meets our criteria, the octet
containing the
N(1710), with a $\delta_h \approx 470$MeV.   However, best
estimates~\cite{PDG} for the partial width for
 N(1710)$\rightarrow \Delta \pi$ is  $\approx 30$MeV, which implies
that the relevant coupling constant is
significantly suppressed
compared to  that in Eq.~(\ref{couplingt}).  We therefore
ignore this multiplet in our subsequent calculations as
it amounts to only a small correction to those of the Roper octet.

For the lowest lying $3/2^-$ octet, we obtain directly from the
experimentally measured
masses~\cite{PDG} that
\footnote{As a self--consistent test of the assignment of
hadrons to this multiplet, note that the violation to the
corresponding GMO relation for this octet is only $15$MeV.}
\begin{equation}
\delta_{8^*} = 296{\rm MeV}.
\end{equation}
The interaction with the $\Delta$ decuplet is, to leading order in the
chiral lagrangian,
\begin{equation}
{\cal L}^i = C^*\left(\overline{T}^\mu \gamma_\nu A^\nu T^*_\mu +
{\rm h. c.}\right).\label{3halfminus}
\end{equation}
The coupling $C^*$  can be determined from the observed decay of the
$N^*(1520)$, $\Gamma(N^*(1520) \rightarrow \Delta \pi) \approx 25$MeV,
whence one finds that
\begin{equation}
\left(\frac{C^*}{\tilde C}\right)^2 \approx \frac{1}{25}.\label{suppC}
\end{equation}
This relative suppression results both from the smaller branching width
 as well as overall kinematic factors that otherwise tend to enhance
$N^*(1520) \rightarrow \Delta \pi$ w.r.t. $N^*(1440) \rightarrow \Delta \pi$.

There are two other $3/2^-$ multiplets, one octet and one decuplet,
 listed in the Particle Data Group that could potentially satisfy our
criteria,  Eq.~(\ref{criteria}).   Each are very poorly determined, containing
 merely one member each, the N(1700) and the $\Delta(1700)$, respectively.
In the case of the N(1700),  its coupling to the $\Delta \pi$ is
experimentally negligible and hence can be safely ignored.
On the other hand, the  coupling with the $\Delta(1700)$ is
not so readily ignored, having a decay width
 $\Gamma(\Delta(1700) \rightarrow \Delta \pi) \approx 120$MeV.
We therefore  explicitly keep this decuplet,
assigning for its intermultiplet spacing with the
$\Delta$ decuplet a value
\begin{equation}
\delta_{10^*} = 1700 - 1232 = 468 {\rm MeV}.
\end{equation}
{}From the aforementioned decay width, and an interaction of the form
Eq.~(\ref{3halfminus}),
we obtain that for the $\Delta^* \Delta \pi$ coupling
\begin{equation}
{\cal H}^{*2} \approx .15.\label{suppH}
\end{equation}
Here we have   used  the convention of Ref.~\cite{Jenkins} for the
$SU_f(3)$ algebra factors
(whereby, for the $\Delta \Delta \pi$ coupling, ${\cal H}^2 \approx 4$
is typical).
As in the case of the $N(1520)$ overall kinematic factors in addition to the
available phase space, yields a rather suppressed value of the coupling.
Indeed, as we will soon see, due to
Eqs.~(\ref{suppC}) and (\ref{suppH}) little
would have been lost had we ignored the  $3/2^-$ multiplets altogether.
 Nevertheless, they have  been included for completeness.

We come finally to the $5/2^+$ states.    The lowest such multiplet is
the N(1680) octet.   It has an intermultiplet mass splitting with the
$\Delta$ decuplet of $\delta_h = 496$MeV.   We conclude that there is
no $5/2^+$ multiplet   that meets our criteria Eq.~(\ref{criteria}).

This then concludes our examination of the relevant multiplets.  By far
the most important,  as we will presently see, is the Roper octet.

\newpage
\section{Decuplet Equal Spacing Rule}
The one--loop, $O(p^3)$ results for the masses of the decuplet
involving intermediate
$\Delta$ decuplet and nucleon octet states have been published
previously~\cite{ourselves}.   The contribution, $\delta M_{10}$,
from the $3/2^-$ multiplet, for the case $m > \delta_{8^*}$, is
\begin{eqnarray}
&\delta M_{10} &= \frac{- 3 \beta \delta_{8^*}}{16 \pi^2 f^2_\pi}
\times\nonumber\\
&&\left[
\left( \delta_{8^*}^2  - \frac12 m^2  \right)
\left( \frac{1}{\epsilon} - \gamma_E +
{\rm ln}(4\pi) + 2- {\rm ln}\frac{m^2 }{\mu^2} \right)
\right. \nonumber\\
&&+\left.
2 \delta_{8^*} \sqrt{m^2 - \delta_{8^*}^2}
\left(\frac{\pi}{2} - {\rm tan}^{-1}\frac{\delta_{8^*}}
{ \sqrt{m^2-\delta_{8^*}^2} }\right)
\right]
\label{uvterms}
\end{eqnarray}
 where $\beta$ represents SU(3) algebra factors~\cite{Jenkins,ourselves}.
As discussed in Ref.~\cite{ourselves}, the counterterms neccessary
to renormalize these terms are either of the
form $\delta_{8^*}^2 {\cal L}_0^{\pi N}$ (for the $\delta_{8^*}^3$ divergences)
or $\delta_{8^*} {\cal L}_1^{\pi N}$ (for the $\delta_{8^*} m^2$ divergences).
As the DES rule is exact for all terms through $m^2$,
all counterterms (divergences) cancel at this order of the chiral expansion
in the violation to the DES rule.

Including all multiplets, the one--loop, ${\cal O}(p^3)$,
violation to the DES rule is
\begin{eqnarray}
(M_{\Sigma^*} &-& M_{\Delta}) - (M_{\Xi^*}-M_{\Sigma^*})
=\nonumber\\
(M_{\Xi^*} &-& M_{\Sigma^*}) - (M_{\Omega^-}-M_{\Xi^*})
=\nonumber\\
\frac12\{(M_{\Sigma^*} &-& M_{\Delta}) -(M_{\Omega^-}-M_{\Xi^*})\}
=\nonumber \\
&&\frac29\, \left( {\cal C}^2 V(-\delta_N) + {\tilde C}^2 V(\delta_R) +
(C^*)^2 V^*( \delta_{8^*}) \right)\nonumber\\
 &&- \frac{20}{81}\, \left( {\cal H}^2 V(0) +
{\cal H^*}^2 V^*( \delta_{10^*}) \right).
\label{DESf}
\end{eqnarray}
\footnote{Note that unlike our convention in ~\cite{ourselves}, all
$\delta_h$ are now strictly positive and hence the explicit
sign in the function $V$ above.}
$V$ and $V^*$ are given by
\begin{eqnarray}
V(\delta) &=& W(m_K,\delta,\mu)\nonumber\\
 &&-\frac34\,W(m_\eta,\delta,\mu)
-\frac14\,W(m_\pi,\delta, \mu),\nonumber\\
V^*(\delta) &=& W^*(m_K,\delta,\mu)\nonumber\\
&&-\frac34\,W^*(m_\eta,\delta,\mu)
-\frac14\,W^*(m_\pi,\delta, \mu),
\label{Vequation}
\end{eqnarray}
wherein the function $W(m,\delta,\mu)$ is~\cite{JenkManconf,BerMeis1}
\footnote{Note the correction from ~\cite{ourselves} regarding the
arctangent term in the case $m>\mid\delta\mid$.}
\begin{eqnarray}
\delta=0,\,\,\,\, && W(m,\delta, \mu)
=\frac{1}{16 \pi f^2_\pi}m^3, \label{del0}\\
m>\mid\delta\mid,\,\,\,\, && W(m,\delta,\mu)
=\nonumber\\ &&\frac{1}{8 \pi^2 f^2_\pi}(m^2 - \delta^2)^{3/2}
\left(\frac{\pi}{2} - {\rm tan}^{-1}\frac{\delta}{ \sqrt{m^2-\delta^2}}\right)
\nonumber\\
&&- \frac{3\delta}{32 \pi^2
f^2_\pi}\left(m^2-\frac{2}{3}\delta^2\right)
{\rm ln}\frac{m^2}{\mu^2}\nonumber\\
\mid\delta\mid>m,\,\,\,\, &&W(m,\delta,\mu) =\nonumber\\
&&\frac{-1}{16 \pi^2f^2_\pi}(\delta^2-m^2)^{3/2}
{\rm ln}\frac{\delta-\sqrt{\delta^2-m^2}}{
\delta+\sqrt{\delta^2-m^2}}\nonumber\\
&&- \frac{3\delta}{32 \pi^2
f^2_\pi}\left(m^2-\frac{2}{3}\delta^2\right)
{\rm ln}\frac{m^2}{\mu^2}\nonumber\\
\end{eqnarray}
and $W^*(m,\delta,\mu)$ is
\begin{eqnarray}
m>\mid\delta\mid,\,\,\,\, && W^*(m,\delta,\mu)
=\nonumber\\ &&\frac{3 \delta^2}{8 \pi^2 f^2_\pi}\sqrt{m^2 - \delta^2}
\left(\frac{\pi}{2} - {\rm tan}^{-1}\frac{\delta}{ \sqrt{m^2-\delta^2}}\right)
\nonumber\\
&&-\frac{3\delta}{16 \pi^2f^2_\pi}( \delta^2-\frac12m^2 )
{\rm ln}\frac{m^2}{\mu^2}\nonumber\\
\mid\delta\mid>m,\,\,\,\, && W^*(m,\delta, \mu)=\nonumber\\
&&\frac{3\delta^2}{16 \pi^2f^2_\pi}\sqrt{\delta^2-m^2}
{\rm ln}\frac{\delta-\sqrt{\delta^2-m^2}}{
\delta+\sqrt{\delta^2-m^2}}\nonumber\\
&&-\frac{3\delta}{16 \pi^2f^2_\pi}( \delta^2-\frac12m^2 )
{\rm ln}\frac{m^2}{\mu^2}.
\end{eqnarray}

As was already mentioned in Section (II),
the contribution from the $3/2^-$ multiplets is
essentially negligible due to their suppressed coupling constants,
Eqs.~(\ref{suppC}) and (\ref{suppH}).    Explicitly we find that,
\begin{eqnarray}
\frac29 C^{*\,2} V^*(+\delta_{8^*}) &\approx& -.06{\rm MeV},
\nonumber\\
-\frac{20}{81}{\cal H}^{*2} V^*(+\delta_{10^*})
&\approx& +.3{\rm MeV} \label{m3/2}
\end{eqnarray}
which are indeed negligible. Hence we omit further considerations of
the $3/2^-$ multiplets.

This is not true of the Roper octet.  Evaluating,
one finds that the ratio of the Roper to Nucleon multiplet contribution
\begin{equation}
 \frac {{\tilde C}^2 V(+\delta_R)} {{\cal C}^2 V(-\delta_N)} = .21
\end{equation}
where result (\ref{couplingt}) has been used.
Taking ${\cal C}^2 = 2$ one obtains that the Roper contribution
\begin{equation}
\frac29{\tilde C}^2 V(+\delta_R) = -5.7{\rm MeV}, \label{mR}
\end{equation}
which is, in absolute magnitude, as large as the average, experimental
value of $6.8$MeV.  It clearly cannot be ignored.

\newpage
\section{Decuplet Magnetic Moments}
The topic of the magnetic moments of the decuplet in the context of chiral
perturbation theory was first discussed in the work of Ref.~\cite{bss}.
Apart from the inclusion of the Roper as an intermediate state, our work
differs from Ref.~\cite{bss} in the treatment of $SU_f(3)$ symmetry.
In our calculation,  the symmetry of the decuplet states
is broken through the meson masses appearing in the one loop calculation.
The meson masses are taken to be proportional  to the current quark masses
with the up and down quark masses being equal.   The strangeness~\cite{Jenkins}
and charge dependences~\cite{LL} of the baryon masses are regarded as effects
of chiral power 1 or more. The quantity $f_K-f_\pi$ has chiral singularity
$\sim m_\pi^2log m_\pi^2$~\cite{DGH} and has chiral power $2$.
Our calculations of the decuplet mass splittings and magnetic moments are
limited to chiral power ${\cal O}(p^3)$ and ${\cal O}(p^2)$, respectively.
Hence, we set $f_K=f_\pi$ and do not include in one loop calculation the sigma
terms from ${\cal L}_1$   which produce strangeness   dependence of baryon
masses at the tree level. We ignore charge-dependence of the masses altogether.
The advantage of this strategy in the calculation of baryon masses is
well-known~\cite{Jenkins,ourselves}. The counter-terms which appear at  one
loop level simply renormalizes the sigma terms. We find a similar result in the
magnetic moment calculation at one loop level, namely, that the counter terms
are strictly proportional to the baryon charge  and hence renormalize
the tree level decuplet magnetic moment term, Eq.~(\ref{treemag}) below.
These advantages are lost if $f_K$ is not set equal to $f_\pi$\cite{bss}.

The lowest order term in the chiral lagrangian for the
magnetic moment of the $i$th member of the $\Delta$
decuplet is given by~\cite{gangof4}
\begin{equation}
{\cal L}_M = - \imath \frac{e}{M_N}
\mu_c q_i \overline{T}^\mu_i T^\nu_i F_{\mu \nu}
\label{treemag}
\end{equation}
where $q_i$ is the charge of the $i$th member.
The one--loop, ${\cal O}(p^2)$ corrections to
the magnetic moments result  from vertex corrections in which the
external photon attaches
to the meson propagator~\cite{bss}  and receives
contributions from intermediate states with
either a $3/2^+$ or $1/2^+$ baryon.
Photon attachments to the intermediate baryon are $m_\pi/M_N$
further suppressed as are also the contributions from $3/2^-$ baryons.
These latter are hence   ignored as they
form part of the higher--order contribution in the chiral expansion.
Note that the $\eta$ meson, being electrically neutral,
also does not contribute
at the order being considered.

Following the notation of ~\cite{gangof4}, the magnetic
moment of the decuplet members, $\mu^{10}_i$, at the ${\cal O}(p^2)$,
one--loop contribution in the chiral expansion
 is, in nuclear magneton units ($e/2 M_N$):
\begin{eqnarray}
 &&\mu^{10}_i = q_i \mu_c + \sum_{ j=\pi,K}\frac{M_N}{32 \pi^2 f_\pi^2}
\left(\alpha^i_j  \, \frac49 {\cal H}^2  {\cal F}(0,m_j,\mu) \right.
\nonumber\\
&&  \left. -\beta^i_j\left[  {\cal C}^2
{\cal F}(-\delta_N,m_j,\mu)
+   {\tilde C}^2 {\cal F}(\delta_R,m_j,\mu) \, \right] \right).
\label{mags}
\end{eqnarray}

The function ${\cal F}(\delta,m,\mu)$ is ultraviolet divergent and given by
\begin{eqnarray}
m>\mid\delta\mid,\,\,\,\, &&{\cal F}(\delta,m,\mu) = \nonumber\\
&&\delta \left( -\frac{1}{\epsilon} + \gamma_E - {\rm ln}(4\pi) - \frac43 +
{\rm ln}(\frac{m^2}{\mu^2})\right)
\nonumber\\
&& - 2 \sqrt{m^2-\delta^2}
\left(\frac{\pi}{2} - {\rm tan}^{-1}\frac{\delta}{ \sqrt{m^2-\delta^2}}\right)
\nonumber\\
\mid\delta\mid>m,\,\,\,\, &&{\cal F}(\delta,m,\mu) = \nonumber\\
&& \delta \left( -\frac{1}{\epsilon} + \gamma_E - {\rm ln}(4\pi) +
{\rm ln}(\frac{m^2}{\mu^2})\right)
\nonumber\\
&&+ \sqrt{\delta^2 - m^2}
{\rm ln}\frac{\delta+\sqrt{\delta^2-m^2}}{\delta-\sqrt{\delta^2-m^2}}.
\label{Fdef}
\end{eqnarray}
The expression for the nonanalytic terms in ${\cal F}(\delta,m,\mu)$
  appeared in Ref.~\cite{bss}.

The   coefficients $\alpha^i_j$ and $\beta^i_j$ are simply related to the
coefficients $\alpha_{ij}$ and $\beta_{ij}$ of Ref.~\cite{bss}.
We multiply the coefficients $\alpha_{ij}$ by $3$ so that
they add up to the charge of decuplet $i$.
Unlike Ref.~\cite{bss}, we use the same mass for all members of a
baryon multiplet.
Accordingly we add the contributions of $\pi^\pm$ and of $K^\pm$.
The sum over $j$ in Eq.~(\ref{mags}) runs over two terms -- $\pi$ and K.
The resulting coefficients have   surprizing  simplicity.  First,
$\alpha^i_j=\beta^i_j$.
Second, they may be expressed  in terms of any two of the following
three -- charge,
$q^i$, isospin, $I^i_3$,  and hypercharge $Y^i$ of decuplet $i$.
All three are traceless in any $SU(3)$ multiplet space.
We choose to use charge and isospin.
\begin{eqnarray}
\alpha^i_\pi=\beta^i_\pi&=&\frac23 I^i_3, \label {ab1} \\
\alpha^i_K=\beta^i_K&=&-\frac23 I^i_3+q^i, \label {ab2} \\
 \alpha^i_\pi+\alpha^i_K&=& \beta^i_\pi+\beta^i_K=q^i. \label{ab3}
\end{eqnarray}

Eq.~(\ref{ab3}) is the key result for renormalizability.  As a
consequence of these relations, the counterterm for the
ultraviolet divergences in ${\cal F}(\delta,m,\mu)$
(which are $m$ independent) is simply proportional to $\delta{\cal L}_M$,
Eq.~(\ref{treemag}), and therefore  absorbed into a redefinition of $\mu_c$.
Note that this is precisely the same procedure as for the one--loop,
$\delta$--dependent contributions to the masses.    We emphasize that this
procedure, and hence renormalizability, is tightly wedded to the systematics of
the chiral expansion (whereby $\delta$ and $f_\pi$ have
fixed values in Eq. (\ref{mags})).

The simplicities of the coefficients $\alpha^i_j$ and $\beta^i_j$
allows great simplification of Eq.~(\ref{mags}).
We introduce the combination:
\begin{eqnarray}
&&{\cal G}_j=\frac{M_N}{32 \pi^2 f_\pi^2}\left[
\frac49 {\cal H}^2  {\cal F}(0,m_j,\mu) \right. \nonumber \\
&&  \left. - {\cal C}^2 {\cal F}(-\delta_N,m_j,\mu)
- {\tilde C}^2{\cal F}(\delta_R,m_j,\mu) \, \right], \label{mags2}
\end{eqnarray}
and rewrite the decuplet magnetic moments in the form:
\begin{equation}
\mu^{10}_i=q_i(\mu_c+{\cal G}_K)+\frac23\,I^i_3({\cal G}_\pi-{\cal G}_K).
\label{mags3} \end{equation}
Note that the   form of Eq.~(\ref{mags3}), in particular the modification of
the coefficient of charge,  reflects the choice of charge and isospin as the
two traceless quantities. The form would be different if we chose charge and
hypercharge or some other combination of isospin and hypercharge.
The second term is present only because the $SU(3)$ symmetry of the states is
broken through the difference in $\pi$ and $K$ masses. If we had used the same
  masses the states would be pure decuplets and the Wigner-Eckart theorem
for $SU_f(3)$ would ensure that the magnetic moments are simply proportional
to the charge.

At  tree level the decuplet magnetic moments are proportional to the
charges only.
The main  one--loop results is the appearance of  the second term of
Eq.~(\ref{mags3}).   Now we need two magnetic moments to fix the two
coefficients in Eq.~(\ref{mags3}), viz,
$\mu_c+{\cal G}_K$ and ${\cal G}_\pi - {\cal G}_K$.
The only decuplet magnetic moment which has been measured is
$\mu_{\Omega^-} = -1.94 \pm .17 \pm .14$ nbm ~\cite{PDG}.
It fixes the coefficient of charge in Eq.~(\ref{mags3}).   For the other
magnetic
moment we choose $\mu_{\Delta^0}$.\footnote{The $\Omega^-$, decaying only
weakly,  is sufficiently long lived to  allow such measurements.
Since all other members of the decuplet decay through the
strong interaction, it is a challenge to extract their magnetic moments
from experiment.}  While not measured yet, it is given entirely by the loop
effect.
\begin{equation}
\mu_{\Omega^-}=-\mu_c-{\cal G}_K,\,\,\,\,\,\mu_{\Delta^0} =
\frac13\,({\cal G}_K - {\cal G}_\pi). \label{mags4}
\end{equation}

We can express the magnetic moments of all other decuplets in terms of these
two magnetic moments.  Specifically, we derive the new relation that
\begin{equation}
\mu^{10}_i=-q_i\,\mu_{\Omega^-} - 2\,I^i_3\,\mu_{\Delta^0}.
\label{mags5}
\end{equation}
Explicit predictions for the eight other decuplet magnetic moments at the
one-loop ${\cal O}(p^2)$ level are listed below.
\begin{eqnarray}
\mu_{\Delta^{++}} &=& - 2 \mu_{\Omega^-} - 3 \mu_{\Delta^{0}}\nonumber\\
\mu_{\Delta^{+}} &=& - \mu_{\Omega^-} -  \mu_{\Delta^{0}}\nonumber\\
\mu_{\Delta^{-}} &=& \mu_{\Omega^-} + 3 \mu_{\Delta^{0}}\nonumber\\
\mu_{\Sigma^{*+}} &=&-  \mu_{\Omega^-} -  2 \mu_{\Delta^{0}}\nonumber\\
\mu_{\Sigma^{*-}} &=&  \mu_{\Omega^-} +  2 \mu_{\Delta^{0}}\nonumber\\
\mu_{\Xi^{*-}} &=&  \mu_{\Omega^-} + \mu_{\Delta^{0}}\nonumber\\
\mu_{\Xi^0} &=&- \mu_{\Delta^{0}}\nonumber\\
\mu_{\Sigma^{*0}} &=& 0.\label{murelate}
\end{eqnarray}
Independent of the explicit mulitplets included as intermediate states,
violations to these relations are {\it strictly} due to higher--order terms
in the chiral expansion.

Note that the magnetic moment of the $\Sigma^{*0}$
continues to be zero at this order in the expansion.    Analogous relations
follow for the quadropole moments\cite{comment}.  We note that
these relations are not obeyed by quenched lattice QCD\cite{Derek}.
This last result is perhaps not surprising as the quenched calculations
do not contain disconnected, quark loop diagrams\cite{Derek2}.

The explicit expression for $\mu_{\Delta^0}$ in terms of the functions
${\cal F}(\delta, m, \mu)$ is given below.
\begin{eqnarray}
\mu_{\Delta^0} =&& \frac{M_N}{96\pi^2 f_\pi^2}
\left( {\cal H}^2 \frac{4}{9}
\left( {\cal F}(0,m_K,\mu) - {\cal F}(0,m_\pi,\mu) \right) \right.\nonumber\\
&&-{\cal C}^2  \left( {\cal F}(-\delta_N,m_K,\mu) -
{\cal F}(-\delta_N,m_\pi,\mu)
\right) \nonumber\\
&&\left.-{\tilde C}^2  \left( {\cal F}(\delta_R,m_K,\mu) -
{\cal F}(\delta_R,m_\pi,\mu) \right) \right).\label{deltalong}
\end{eqnarray}
With the help of Eqs.~(\ref{Fdef}) it is easy to verify that
$\mu_{\Delta^{0}}$ is renormalization scale independent.

Since the magnetic moment of the ${\Delta^0}$ is given strictly by
loop--effects, it is an appropriate measure of the relative importance of
the Roper at the one--loop level.  Explicitly one has from
Eq.~(\ref{deltalong}) that
\begin{equation}
 \mu_{\Delta^{0}} = -.055 {\cal H}^2 + .229 {\cal C}^2 + .086 {\tilde C}^2,
\label{deltanumber}
\end{equation}
As in the case of the masses\cite{Jenkins} there is a strong
cancellation between the $\Delta$ decuplet and nucleon
octet intermediate states.    This implies that $\mu_{\Delta^0}$
is a very delicate  function of ${\cal H}^2$ and $ {\cal C}^2$
and therefore potentially very sensitive to the Roper contribution
(${\tilde C}^2$).      Ambiguity in this regard resides in the fact
that ${\cal H}^2$ and $ {\cal C}^2$ are not sufficiently well--known
that  a reliable prediction of $\mu_{\Delta^{0}}$ can be made.
To illustrate this point, we quote the results using
two  ``representative''  values for the couplings, both with and without
the Roper included.   For the coupling values used in Ref. \cite{ourselves},
one obtains only a mild dependence on the Roper,
 \begin{eqnarray}
\mu_{\Delta^{0}} ( {\cal H}^2 = 4.4, {\cal C}^2 = 2,
{\tilde C}^2 = 0) &=& .21,\nonumber\\
\mu_{\Delta^{0}} ( {\cal H}^2 = 4.4, {\cal C}^2 = 2,
{\tilde C}^2 = 1) &=& .29,
\end{eqnarray}
while for the couplings used by Ref. {\cite{bss}}, we find a
much more dramatic dependence on the Roper,
\begin{eqnarray}
\mu_{\Delta^{0}} ( {\cal H}^2 = 4.84, {\cal C}^2 = 1.44,
{\tilde C}^2 = 0.0) &=& .05,\nonumber\\
\mu_{\Delta^{0}} ( {\cal H}^2 = 4.84, {\cal C}^2 = 1.44,
{\tilde C}^2 = .72) &=& .12.
\end{eqnarray}
The difference in $\mu_{\Delta^{0}}$ from these two parameter sets
is clearly sizable, as is the relative importance of the Roper.

If instead we chose to use as input the recent, model dependent
extraction\cite{experiment2} of the magnetic moment of the $\Delta^{++}, \,
\mu_{\Delta^{++}} = 4.5 \pm .5$ to infer  $\mu_{\Delta^{0}}$ using the
relations Eq.~(\ref{murelate}),  one then obtains that
$\mu_{\Delta^{0}} = -.2 \pm .2$.     If this is indeed the data, then by
Eq.~(\ref{deltanumber}) the contribution of the Roper is, in absolute
magnitude, significant.    As in the case of the mass--splittings,
a formulation of $\chi$PT without the Roper as  an explicit degree of
freedom is intrinsically incapable of predicting such ``data''.

\newpage
\section{Conclusions}

{}From the results of the last two sections a few points are worth discussing.

First,  that while the magnetic moment of the $\Delta^0$ sensitively depends
on the cancellation between terms depending on relatively ill--determined
coupling constants,  the relations between the magnetic moments of the $\Delta$
decuplet given in Eq. (\ref{murelate}) are rigorous predictions of $\chi$PT
at $O(p^2)$.
We urge experimental activity to confront these predictions with
data. A  new measurement  of  $\mu_{\Omega^-}$ with higher precision
will be most useful. At least two other decuplet magnetic moments need
to be measured, hopefully with a precision of $\sim 0.1$ n.m..

Second, that on the level of one--loop corrections in $\chi$PT,  the
contribution of
the Roper octet to properties of the $\Delta$ decuplet is as important
as any other  multiplet's contribution.    We have   seen this result
quantitatively in the case of the DES rule and the magnetic moment of
the $\Delta^0$.      Both of these   quantities are good measures of the
one--loop effects as they  are each zero  at lower--order in the chiral
expansion.
We expect that these results are illustrative and that they generalize to all
one--loop calculations for the $\Delta$ decuplet.
Since the mass--splitting, $\delta_R$, between the Roper octet and
$\Delta$ decuplet is less than that of the kaon's mass,  a Taylor expansion
in $m_K/\delta_R$ is not permissible.   Hence, these loop
effects cannot be  be absorbed within higher order
terms of the chiral expansion of a theory not containing
the Roper as an explicit degree of freedom.
In such a theory it is indeed difficult to justify going to one--loop
or higher without inclusion of the Roper.   Phenomenologically successful
results would have to be considered merely fortuitous unless shown
to be a result of more general considerations
(as in the relations of Eq. (\ref{murelate}) for the magnetic moments).

Third, that the above argument can be repeated in kind for the one--loop
corrections to the Roper resonance.      That is, even higher--multiplets,
separated  by $\delta_h$ in mass from the Roper octet by an
amount less than the kaon mass, will {\it a priori} be as important
quantitatively as the $\Delta$ decuplet for
properties of the Roper at the one--loop level.  Since such corrections
are neccessary for a two loop calculation
\footnote{See though ~\cite{ourselves} for a discussion as to the
likely feasibility of such a program.} of the baryon masses,
we are led to conclude that the loop expansion in
general in the baryon sector of $\chi$PT is inevitably wedded to the
neccessity of including more and more multiplets in the theory
as fundamental fields.     While such a result may not
be true in a particular limit of QCD
({\it e.g.} $m_{u, d, s} \rightarrow 0$ or $N_c \rightarrow \infty$),
it is a consequence of the experimental fact
that on the average, $m_\pi \approx \delta_h$.

\bigskip
{\centerline{ACKNOWLEDGEMENTS}} We thank
  Y. Umino and T. D. Cohen for useful discussions.
This work was supported in part by DOE Grant DOE-FG02-93ER-40762.


\end{document}